\def\ref{\par\noindent\hang}

\def\spose#1{\hbox to 0pt{#1\hss}}
\def\approxlt{\mathrel{\spose{\lower 3pt\hbox{$\sim$}}
        \raise 2.0pt\hbox{$<$}}}
\def\approxgt{\mathrel{\spose{\lower 3pt\hbox{$\sim$}}
        \raise 2.0pt\hbox{$>$}}}

\def\multleft#1{\hbox to size{\vbox {\halign {\lft{##}\cr #1}}\hfill}\par}
\def\multright#1{\hbox to size{\vbox {\halign {\rt{##}\cr #1}}\hfill}\par}

\def\degmark{^\circ}
\def\today{\ifcase\month\or January\or February\or March\or April\or May\or
      June\or July\or August\or September\or October\or November\or December\fi
      \space\number\day, \number\year}
\def\$<${\thinspace}

\def\boxit#1{\vbox{\hrule\hbox{\vrule\kern3pt\vbox{\kern3pt
          #1 \kern3pt}\kern3pt\vrule}\hrule}}

\def\eV{{\rm\thinspace eV}}

\def\keV{{\rm\thinspace keV}}

\documentstyle[psfig]{mn}

\title[Special relativistic effects on iron line strength.]
{Special relativistic effects on the strength of the fluorescent K$\alpha$
iron line from black hole accretion disks.}
\author[C.~S.~Reynolds \& A.~C.~Fabian]
{C.~S.~Reynolds$^1$ and A.~C.~Fabian$^2$\\
{\small $^1$JILA, University of Colorado, Boulder, Colorado, CO~80309-0440,
USA}\\
{\small $^2$Institute of Astronomy, Madingley Road, Cambridge, CB3 0HA}\\
}
 
\begin{document}
 
\maketitle
 
\begin{abstract}
The broad iron K$\alpha$ emission line, commonly seen in the X-ray spectrum
of Seyfert nuclei, is thought to originate when the inner accretion disk is
illuminated by an active disk-corona.  We show that relative motion between
the disk and the X-ray emitting material can have an important influence on
the observed equivalent width (EW) of this line via special relativistic
aberration and Doppler effects.  We suggest this may be relevant to
understanding why the observed EW often exceeds the prediction of the
standard X-ray reflection model.  Several observational tests are suggested
that could disentangle these special relativistic effects from iron
abundance effects.
\end{abstract}


\section{Introduction}

Recent X-ray observations of many active galactic nuclei (AGN) have found
the K$\alpha$ emission line of cold iron (at 6.40\,keV), and show this line
to be broadened and skewed towards low-energies (Mushotzky et al. 1995;
Tanaka et al. 1995; Nandra et al. 1997).  The line profile is usually found
to be in good agreement with that expected if it were to originate from the
inner regions of an accretion disk around a black hole, with gravitational
redshifts and (relativistic) Doppler shifts being the processes relevant to
shaping this profile.  In the cases with the best data, many alternative
mechanisms for forming a such a line can be examined and rejected (Fabian
et al. 1995).

The iron line is believed to be a fluorescent line which results when
Thomson-thick material, with a relatively low ionization state, is
externally illuminated by a hard X-ray source (Basko 1978; George \& Fabian
1991; Matt, Perola \& Piro 1991).  In the case of accreting black hole
systems, the optically-thick material can be identified with a thin
accretion disk whereas the external X-ray source is probably a disk-corona.
The equivalent width (EW), $W$, of the iron line can be predicted given the
source geometry, illuminating X-ray spectrum, and chemical abundances.  In
the `standard' case where the cold material possesses Morrison \& McCammon
(1983) cosmic abundances and subtends $2\pi$\,sr at the X-ray source (which
is assumed to have an AGN-like spectrum), the EW is $W_{\rm mm}=150\eV$.
For a more recent set of cosmic abundances (Anders \& Grevesse 1993), this
EW increases slightly to $W_{\rm ag}=190\eV$ (Reynolds, Fabian \& Inoue
1995; Matt, Fabian \& Reynolds 1997), primarily due to the increased
abundance of iron.

Observationally, many iron lines are significantly stronger than the
predictions of the previous paragraph.  Although the mean EW in the sample
of Nandra et al. (1997) is only slightly in excess of $W_{\rm ag}$,
$\langle W \rangle = 230\eV$, there are many objects in this sample with
very strong lines, $W\sim 300-600\eV$.  Several possible explanations for
the strength of these lines have been previously discussed.  First, the
iron may be over-abundant (George \& Fabian 1991; Reynolds, Fabian \& Inoue
1995).  However, the EW grows only logarithmically with iron abundance due
to the fact that iron itself contributes to its own K$\alpha$ line opacity
through L-shell photoelectric absorption.  Thus, extreme iron
overabundances ($\sim 10\,{\rm Z}_\odot$ or more) are required to explain
the strongest lines.  Secondly, the geometry might be such that the cold
material subtends more than $2\pi$\,sr at the X-ray source.  This geometry
is difficult to reconcile with a disk/corona model.  Thirdly, gravitational
focusing of X-ray flux from a source which is at some height above the disk
plane can enhance the EW of the line (Martocchia \& Matt 1996; Reynolds \&
Begelman 1997).  However, these General Relativistic (GR) effects are only
important for enhancing the emission from the innermost regions of the disk
(radii $r<3\,{\rm R}_{\rm Sch}$, where ${\rm R}_{\rm Sch}$ is the
Schwarzschild radius of the central black hole), which produce very
redshifted line emission (with observed photon energies of $\sim 4\keV$ or
less).  Thus, whilst GR enhancement effects might be important for
understanding the broadest line known, they cannot be relevant to strong
iron lines from typical objects.

In this {\it letter}, we note that any relative motion between the X-ray
source and the accretion disk will also affect (and usually enhance) the EW
of the iron line through the effects of special relativistic (SR)
aberration and Doppler shifts.  Such relative disk/corona motion will
naturally occur if the accretion disk and the corona are not rigidly
coupled together.  For example, some authors treat the disk-corona as an
independent, slim accretion disk.  Due to the subsequent sub-Keplerian
motion of the corona, the disk and corona will be in relative motion at any
given radius.  However, the corona may well be tightly coupled to the
accretion disk by magnetic fields (which will force the disk and corona to
accrete together).  Even in this circumstance, the SR
effects discussed here may be important.  Field \& Rogers (1993; hereafter
FR93) have argued that magnetic instabilities and reconnection events in a
disk corona could produce shock waves and/or the streaming of relativistic
particles along the magnetic field lines.  Thus, the plasma which is
instantaneously responsible for the X-ray emission might well be in bulk
motion relative to the disk.  Such arguments gain qualitative support by
drawing an analogy with the solar corona and solar flares.

For convenience, we set the speed of light to unity, $c=1$, throughout this
work.

\section{A stationary slab illuminated by a moving source}

Most of the SR effects relevant to the EW of the iron
line can be studied within a scenario in which the iron line originates
from a stationary slab of cold matter.

Suppose we have a semi-infinite slab of cold gas filling the half-space
$z<0$.  Let this cold slab be stationary with respect to a distant observer
who views it at an inclination $i$ with respect to the upward normal to
the slab.  Furthermore, suppose this cold slab is illuminated by an X-ray
source which is at some distance $h$ above the face of the slab.  We make
the following assumptions about the X-ray source:
\begin{enumerate}
\item the source is pointlike.
\item the X-ray emission is isotropic in the rest-frame of the source.
\item the spectrum of the emission, when viewed in the rest frame, is a 
power-law with photon index $\Gamma=2$ for all energies relevant to this
discussion.  This choice of photon index is observationally motivated
(e.g. Nandra \& Pounds 1994; Reynolds 1997).
\end{enumerate}
We further assume that the source is moving at velocity $v$ relative to
the slab.  We define $\alpha$ to be the angle this velocity vector makes
with the downward normal (thus $\alpha=0$ corresponds to motion directly
towards the slab).  We also define $\beta$ to be the azimuthal direction of
the source in the slab plane relative to some reference line on the slab.
We choose the projection of the observers line of sight to be this
reference line.  Finally, we construct a standard 2-d polar co-ordinate
system $(r,\phi)$ on the slab taking the point that is (instantaneous)
below the source to be the origin, and using same reference direction used
to define $\beta$.

From simple vector geometry, we can calculate two other important angles.
First, for a given point on the slab $(r,\phi)$, the angle that the
source velocity makes with the line joining the source with that point,
$\theta$, is given by
\begin{equation}
\cos\theta=\frac{r\sin\alpha\cos\beta\cos\phi+r\sin\alpha\sin\beta\sin\phi+h\cos\alpha}{(r^2+h^2)^{1/2}}.
\end{equation}
Secondly, the angle between the observers line of sight and the source
velocity, $\eta$, is
\begin{equation}
\cos\eta=\sin\alpha\cos\beta\sin i-\cos\alpha\cos i.
\end{equation}

The question we wish to address is this: how does the relative slab-source
motion influence the strength of the iron fluorescence line that will
result from the illumination.  There will be two relevant effects.  First,
the Doppler shifts and aberration of the source emission will influence both
the number and average direction of primary photons that fall above the
iron photoelectric threshold.  This will change the {\it absolute} line
photon emission rate as compared with the static source case.  Secondly,
the fact that the primary emission suffers relativistic aberration whereas
the line emission does not will affect the observed ratio of these two
emissions and, hence, the EW of the iron line.   We will now discuss these
two effects in turn.

\subsection{Influence of the motion on the absolute line photon emission
rate}
 
There are two distinct ingredients involved in determining the absolute
emission rate of fluorescent iron line photons.  The most obvious one is
the number of illuminating photons with energy $E>E_{\rm th}$, where
$E_{\rm th}=7.1\keV$ is the photoelectric
threshold for neutral iron.  Only these incident photons can eject one of
the K-shell electrons from iron and, thus, initiate the radiative cascade
within the atom that results in a K$\alpha$ line photon being emitted.  For
a fixed illuminating spectrum, the iron line strength is simply
proportional to the normalization of that spectrum. 

The second ingredient in determining the absolute iron line emission is the
geometry of the illumination (Basko 1978; George \& Fabian 1991).  Consider
normally incident photons with energy $E>E_{\rm th}$.  On average, such
photons traverse a distance corresponding to unity optical depth prior to
being photoelectrically absorbed.  The resulting iron fluorescence photons
have to travel through at least the same optical depth of material in order to
escape the slab.  Some fraction of these iron line photons will be
absorbed in this process (either by K-shell photoionization of
low-$Z$ metals or L-shell photoionization of iron).  Now, consider photons that
are incident on the slab with a large inclination (i.e. grazing
collision).  Again, these photons are absorbed after a unity optical depth,
but this now corresponds to a significantly smaller vertical depth in the
slab.  Thus, the resultant iron line photons can escape significantly
less impeded by subsequent absorption.  The net result is that the {\it
effective fluorescent yield} increases with increasing inclination (see
Fig.~1 of George \& Fabian 1991).

This geometrical effect only becomes important when most of the incident
flux strikes the disk at a high inclination.  In the scenario under
discussion here, that corresponds to irradiation by sources which are
rapidly moving in a direction parallel to the disk plane.  Given this fact,
and that analytical descriptions of the geometrical dependence are somewhat
cumbersome (and approximate; e.g. Basko 1978), we shall ignore this effect
and study just the Doppler-shifting phenomenon.  
 
\begin{figure}
\psfig{figure=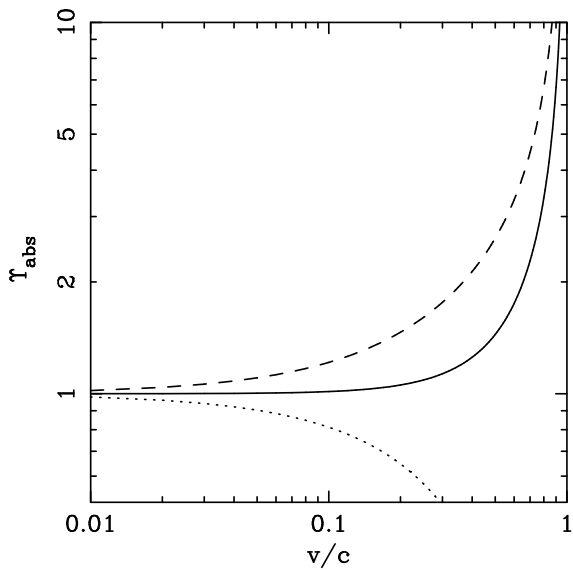,width=0.65\textwidth,angle=270}
\caption{Enhancement in the absolute fluorescent line emission from the
stationary slab as a function of source velocity.  Shown here are the cases
$\alpha=0$ (source motion directly towards slab; dashed line),
$\alpha=\pi/2$ (source motion parallel to the slab; solid line), and
$\alpha=\pi$ (source motion directly away from slab; dotted line).}
\end{figure}

With this restriction, and the assumption of a power-law primary spectrum,
the problem simply amounts to determining how the normalization of the
illuminating spectrum at some given energy ($E=E_{\rm th}$, say),
integrated over the surface of the slab, is affected by the source motion.
Making use of the phase space invariant $I_\nu/\nu^3$, we see that the
SR enhancement in absolute iron line production is
\begin{equation}
\Upsilon_{\rm abs}=\frac{\int\int F(r)
\delta(\theta)^{(2+\Gamma)}\,r\,dr\,d\phi}{\int\int F(r) \,r\,dr\,d\phi},
\end{equation}
where $F(r)$ is the illuminating flux striking a unit area of the slab in
the absence of any relativistic effects, 
\begin{equation}
F(r)\propto\frac{h}{(r^2+h^2)^{3/2}},
\end{equation}
and $\delta(\theta)$ is the {\it beaming parameter}, 
\begin{equation}
\delta(\theta)=\frac{(1-v^2)^{1/2}}{(1-v\cos\theta)}.
\end{equation}
Simple expressions can be obtained for $\Upsilon_{\rm abs}$ in three cases:
\begin{enumerate}
\item source motion directly towards the slab ($\alpha=0$):
\begin{equation}
\Upsilon_{\rm abs}=\frac{(1-v+\frac{1}{3}v^2)(1+v)^2}{1-v},
\end{equation}
\item source motion parallel to the slab ($\alpha=\pi/2$):
\begin{equation}
\Upsilon_{\rm abs}=\frac{1+\frac{1}{3}v^2}{1-v^2},
\end{equation}
\item source motion directly away from the slab ($\alpha=\pi$):
\begin{equation}
\Upsilon_{\rm abs}=\frac{(1+v+\frac{1}{3}v^2)(1-v)^2}{1+v}
\end{equation}
\end{enumerate}

In Fig.~1, we plot $\Upsilon_{\rm abs}(v)$ for these cases.  As intuitively
expected, the former two cases enhance the absolute iron emission.  For
motion along the slab normal ($\alpha=0,\pi$), the beaming can influence
the absolute line production by a factor of two for velocities that are
only mildly relativistic ($v\sim 0.4$).

\subsection{Differential beaming and the equivalent width of the line}
 
The fact that the primary emission is beamed whereas the fluorescent
emission is not beamed has direct consequences for the observed EW of the
line.  The relevant quantity is the ratio of the iron line flux to the
normalization of the observed primary continuum at the iron line energy.
Due to the effects of relativistic beaming, this primary flux normalization
is proportional to $\delta(\eta)^{2+\Gamma}$ where, to recap, $\eta$ is the
angle between the source motion and the observers line of sight.  Noting
that $\Gamma=2$, the equivalent width of the iron line is given by
\begin{equation}
\frac{W(v)}{W(0)}=\frac{\Upsilon_{\rm abs}\left[1-v(\sin\alpha\cos\beta\sin i-\cos\alpha\cos i)\right]^4}{(1-v^2)^2}.
\end{equation}

\begin{figure*}
\hbox{
\psfig{figure=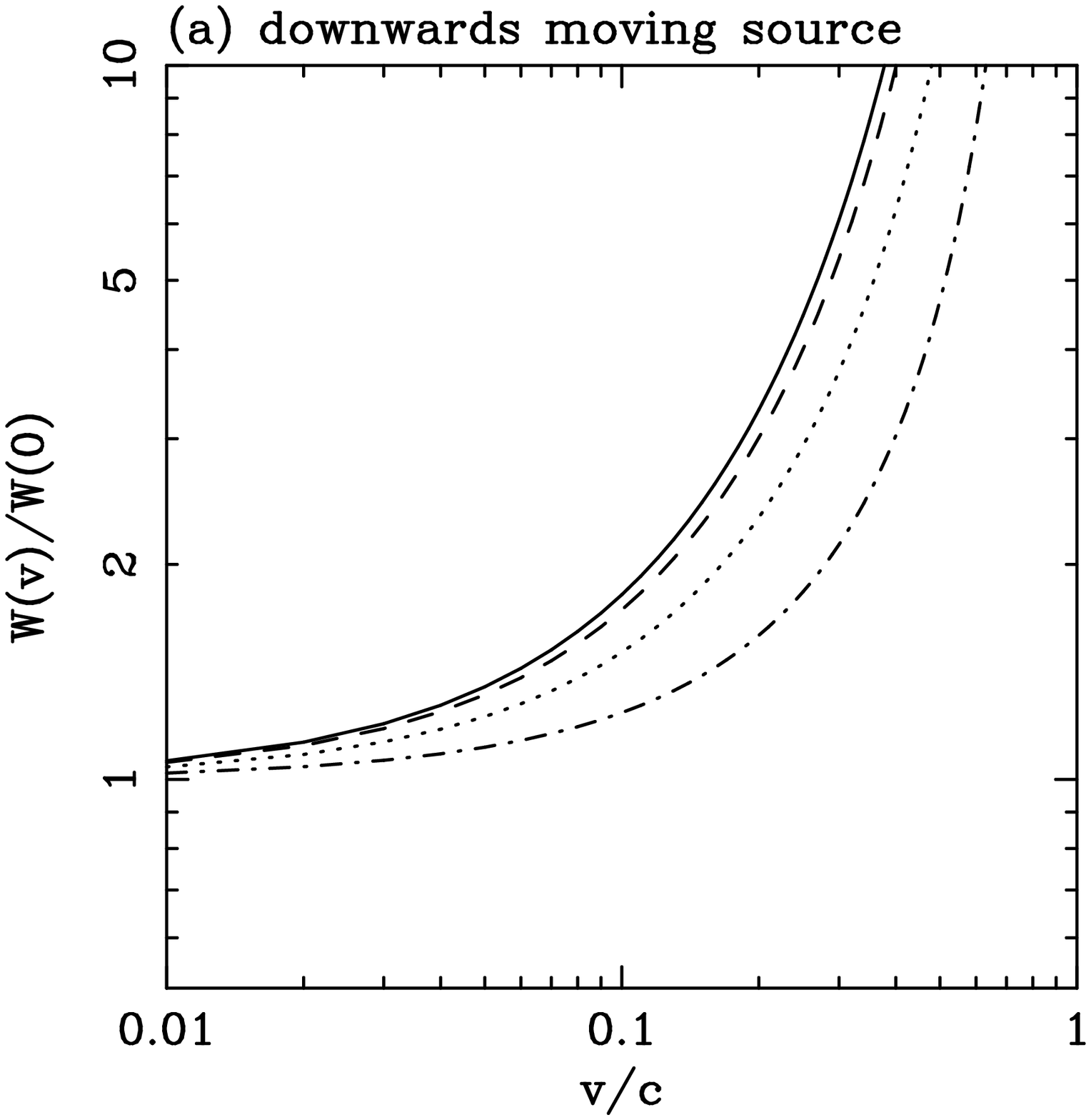,width=0.6\textwidth,angle=270}
\hspace{-2cm}
\psfig{figure=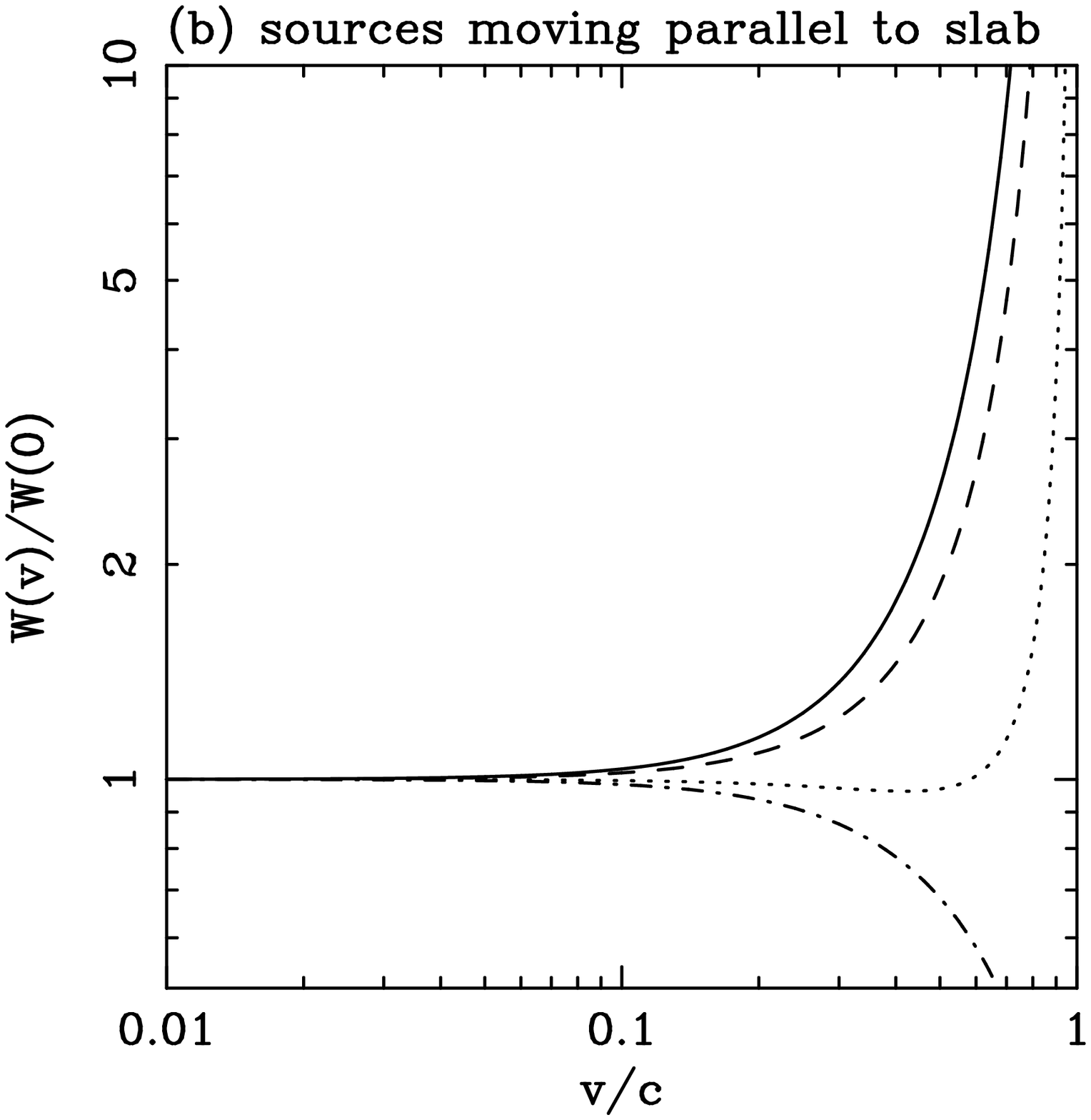,width=0.6\textwidth,angle=270}
}
\caption{The enhancement in the iron line equivalent width $W(v)/W(0)$ for
(a) downwards moving sources with velocity $v$, and (b) an ensemble of
sources moving with speed $v$ (isotropically) parallel to the slab
plane.   Four different inclinations are shown: $i=0\degmark$ (solid line),
$i=30\degmark$ (dashed line), $i=60\degmark$ (dotted line), and
$i=90\degmark$ (dot-dashed line).}
\end{figure*}

Figure~2a shows the behaviour of $W(v)$ for sources moving directly towards
the slab ($\alpha=0$) for various inclinations of the observer $i$.  It can
be seen that the SR beaming has a major effect on the EW of the iron line
for relatively small velocities, especially at low inclinations.  The EWs
seen in Seyfert 1 nuclei (which have typical inclinations of $i\sim
30\degmark$) are often 2--3 times more than predicted by the standard model
(see Introduction).  If all of this enhancement was due to SR effects
(rather than iron overabundance), we would require downwards motion at
$v\sim 0.1-0.2$.  By contrast, Fig.~2b shows the behaviour of $W(v)$ for
sources moving parallel to the plane of the slab ($\alpha=\pi/2$).  In
plotting these curves, it is assumed that there is an ensemble of such
sources with velocity vectors that are isotropically distributed in that
plane (i.e. we average over $\beta$).  Once this average is performed, the
resulting expression for the equivalent width is
\begin{equation}
\frac{W(v)}{W(0)}=\frac{(1+\frac{1}{3}v^2)(1-v^2\sin^2i)^{7/2}}{(1+v)(1-v)^3(1+\frac{3}{2}v^2\sin^2i)}.
\end{equation}
It is seen from Fig.~2b that much higher source velocities are required to
produce a given enhancement in the iron line.  To explain Seyfert 1 spectra
would require $v\sim 0.5$.

\section{Discussion and conclusions}

Using a stationary slab model, we have shown that relative disk/source
motion can significantly affect (and usually enhance) the EW of the
fluorescent iron K$\alpha$ line.  This mechanism might be responsible for
producing the strong iron lines seen in many AGN.  Viable alternative
explanations require a rather large iron overabundance, or invoke some
geometry in which the disk subtends more than $2\pi$\,sr as seen by the
X-ray source.

As is intuitively clear, the SR enhancement of the line
EW is especially important if the X-ray emitting material is moving
directly towards the disk.  This is exactly the flow pattern envisaged in
the magnetized accretion disk model of FR93.  By analogy with solar flares,
they argue that the corona is magnetically confined by loops of magnetic
field which have footpoints in the accretion disk.  They suggest two
processes that might leads to significant material motion.  Firstly, if the
footpoints of the loops are forced (by motions in the disk) into a
configuration whose topology permits a lower energy magnetic field
structure, the coronal loop structure may become unstable, leading to the
release of magnetic field energy into bulk kinetic energy of the plasma.
Coronal shock waves will result.  Secondly, the disk motion may force loops
of opposite polarity together, thereby resulting in magnetic reconnection.
This would also channel a substantial amount of energy into particle
energy.  Since these phenomena are likely to occur predominately near the
tops of the coronal loops (Field \& Rogers 1992), the streaming of
accelerated particles along the field lines will result in downwards motion
of X-ray emitting material.  FR93 use the subsequent beaming of the
emission to channel much of the coronal energy back into the accretion disk
(which is eventually lost as optical/UV thermal emission from the disk
surface).

In the very near future, observations will begin to address the origin of
the strong iron lines.  Very broad band X-ray observations (with {\it RXTE}
or {\it Beppo-SAX}, for example) will allow the iron line, iron edge and
Compton backscattered continuum to be simultaneously constrained.  If the
iron lines are strong due to an iron-overabundance, this will be revealed
as an enhancement of the line relative to the Compton backscattered
continuum.  On the other hand, if observations show both the iron line {\it
and} backscattered continuum to be enhanced above the predictions of the
`standard' model, then we must either consider a geometry in which the disk
subtends more than $2\pi$\,sr at the X-ray source, or invoke (special or
general) relativistic effects to enhance the overall X-ray reflection.  It
is interesting to note that preliminary results for the Seyfert 1 galaxy
MCG$-$6-30-15 indicate that the backscattered continuum may somewhat
enhanced above the prediction of the standard model (Molendi et al. 1997),
although a self-consistent analysis, which includes the effects of iron
abundance on the shape of the backscattered continuum, still has to be
performed.

If it is confirmed that an overall enhancement of the X-ray reflection is
required, the inclination dependence of the iron line properties will be
important for disentangling the enhancement mechanisms.  As shown in this
work, if source motion is responsible for the enhancement, the EW of the
line will decrease significantly as one considers sources at higher
inclination (note that the inclination of the disk can be measured robustly
from the iron line profile).  This effect will be much stronger than the
inclination dependence of the line just based on limb-darkening (George \&
Fabian 1991).  A careful analysis of existing {\it ASCA} datasets might
allow such a trend to be addressed.

\section*{acknowledgments}

CSR gratefully acknowledges support from the National Science Foundation
under grant AST9529175.  ACF thanks the Royal Society for support.

\end{document}